\documentclass[unnumsec,webpdf,contemporary,large]{oup-authoring-template}%

\graphicspath{{figs/}}
\usepackage{longtable}
\usepackage{makecell}
\usepackage{hyperref}

\theoremstyle{thmstyleone}%
\theoremstyle{thmstyletwo}%
\theoremstyle{thmstylethree}%

\begin{document}

\journaltitle{Wei et al., 2024}
\DOI{In Preparation}
\copyrightyear{2024}
\appnotes{Applications Note}
\firstpage{1}

\title{wgatools: an ultrafast toolkit for manipulating whole genome alignments}

\author[1,2,$\#$]{Wenjie Wei\ORCID{0000-0003-4105-9693}}
\author[3,$\#$]{Songtao Gui\ORCID{0000-0001-6718-982X}}
\author[2,4]{Jian Yang\ORCID{0000-0003-2001-2474}}
\author[5]{Erik Garrison\ORCID{0000-0003-3821-631X}}
\author[1]{Jianbing Yan\ORCID{0000-0001-8650-7811}}
\author[6,$\ast$]{Hai-Jun Liu\ORCID{0000-0001-7717-893X}}

\authormark{Wei et al. \textit{wgatools}.}

\address[1]{\orgdiv{National Laboratory of Crop Genetic Improvement}, \orgname{Huazhong Agricultural University}, \postcode{430070}, \state{Wuhan}, \country{China}}
\address[2]{\orgdiv{School of Life Sciences}, \orgname{Westlake University}, \postcode{310030}, \state{Hangzhou}, \country{China}}
\address[3]{\orgdiv{National Key Laboratory of Wheat Improvement, College of Life Sciences}, \orgname{Shandong Agricultural University}, \postcode{271018}, \state{Tai’an}, \country{China}}
\address[4]{\orgdiv{Westlake Laboratory of Life Sciences and Biomedicine}, \postcode{310024}, \state{Hangzhou}, \country{China}}
\address[5]{\orgdiv{Department of Genetics, Genomics and Informatics}, \orgname{University of Tennessee Health Science Center}, \postcode{38163}, \state{Memphis, TN}, \country{USA}}
\address[6]{\orgdiv{Yazhouwan National Laboratory}, \postcode{572024}, \state{Sanya}, \country{China}.}
\corresp[$\#$]{These authors contributed equally to this work.}
\corresp[$\ast$]{Corresponding authors: Hai-Jun Liu (\href{mailto:liuhaijun@yzwlab.cn}{liuhaijun@yzwlab.cn}).}


\abstract{
\textbf{Summary:}
With the rapid development of long-read sequencing technologies, the era of individual complete genomes is approaching. We have developed \textit{wgatools}, a cross-platform, ultrafast toolkit that supports a range of whole genome alignment (WGA) formats, offering practical tools for conversion, processing, statistical evaluation, and visualization of alignments, thereby facilitating population-level genome analysis and advancing functional and evolutionary genomics.\\
\textbf{Availability and Implementation:} \textit{wgatools} supports diverse formats and can process, filter, and statistically evaluate alignments, perform alignment-based variant calling, and visualize alignments both locally and genome-wide. Built with Rust for efficiency and safe memory usage, it ensures fast performance and can handle large datasets consisting of hundreds of genomes. \textit{wgatools} is published as free software under the MIT open-source license, and its source code is freely available at \href{https://github.com/wjwei-handsome/wgatools}{\textit{https://github.com/wjwei-handsome/wgatools}}.\\
\textbf{Contact:} \href{mailto:weiwenjie@westlake.edu.cn}{weiwenjie@westlake.edu.cn} (W.W.) or \href{mailto:liuhaijun@yzwlab.cn}{liuhaijun@yzwlab.cn} (H.-J.L.).\\
}

\keywords{Long-read sequencing, Whole genome alignment, Variant calling, Data structure, Rust}

\maketitle

\section{Introduction}
The advent of long-reads sequencing technologies has revolutionized genomics, enhancing the continuity and feasibility of sequencing complete genomes \cite{van_dijk_third_2018,li_genome_2024}. This technological advancement is paving the way for an era where personalized genomes could become a common resource for scientific research and medical applications. Whole genome alignment (WGA), a foundational technique in comparative genomics, plays a critical role in the analysis and interpretation of genomic data. It facilitates the identification of genetic variations and evolutionary relationships among different individuals or species.
WGA techniques vary widely, each developed to address specific research needs and to optimize particular aspects of genome analysis \cite{anisimova_whole-genome_2019,kille_multiple_2022,song_new_2024}. These techniques generate data in multiple formats, such as MAF (Multiple Alignment Format, \url{https://genome.ucsc.edu/FAQ/FAQformat.html#format5}), PAF (Pairwise mApping Format, \url{https://github.com/lh3/miniasm/blob/master/PAF.md}) \cite{li_minimap2_2018}, and Chain (\url{https://genome.ucsc.edu/goldenPath/help/chain.html}), which are tailored for distinct analytical purposes (Table \ref{tab:formats}). However, the diversity of these formats poses a significant challenge: incompatibility between them impedes the seamless integration and comparison of genomic data across different studies or platforms. Consequently, researchers often find themselves confined to the data types supported by their chosen tools, which can limit the scope of their analyses and hinder collaborations.

\begin{table*}[b]
\caption{\textbf{Comparison of the widely used genome alignment formats}.
\label{tab:formats}}
\tabcolsep=0pt
\begin{tabular*}{\textwidth}{@{\extracolsep{\fill}}lcccccc@{\extracolsep{\fill}}}
\toprule%
\textbf{Format} & \textbf{Application Scenarios} & \textbf{Structure} & \textbf{Pros} & \textbf{Cons} & \textbf{Type} \\
\midrule

Chain & {\makecell[l]{\\Suitable for large-scale\\ genome assembly and\\ cross-species comparisons;\\ Used to represent syntenic\\ regions.}}  & {\makecell[l]{Links sets of\\ alignment blocks\\ that are homologous\\ and ordered\\ in both genomes.}}  & {\makecell[l]{Useful for long-range\\ relationships and\\ annotation transfer.}}  & {\makecell[l]{Lacks base-pair level\\ detail, focusing more\\ on structure.}}  & Pairwise \\
PAF & {\makecell[l]{Efficient in long-read\\ sequencing for storing\\ large genomic alignments.}}  & {\makecell[l]{Tab-delimited, includes\\ basic alignment data\\ like names, lengths,\\ positions, and mapping\\ quality.}}  & {\makecell[l]{Efficient with large,\\ long-read datasets.}}  & {\makecell[l]{Omission of finer alignment\\ details which may be\\ crucial for certain analyses.}}  & Pairwise \\
MAF & {\makecell[l]{Best for comparative\\ genomics across\\ multiple species,\\ phylogenetics, and\\ evolutionary studies.}}  & {\makecell[l]{Contains blocks with\\ alignments, each block\\ starts with `a' and\\ sequence lines start\\ with `s'.}}  & {\makecell[l]{Excellent for multi-species\\ alignments and\\ detailed analysis.}}  & {\makecell[l]{Bulky and less\\ efficient for very\\ large datasets.}}  & Multiple \\
Delta & {\makecell[l]{Ideal for closely\\ related genomes\\ or small-scale differences;\\ Used by MUMmer\\ for basic differences\\ between sequences.\\}}  & {\makecell[l]{Consists of a header\\ and alignment blocks\\ detailing insertions,\\ deletions, and \\substitutions.}}  & {\makecell[l]{Compact and efficient \\for similar sequences.}}  & {\makecell[l]{Less suitable for\\ complex rearrangements\\ and lacks detailed\\ visualization.}}  & Pairwise \\
\botrule
\end{tabular*}
\end{table*}

Recognizing these challenges, there is a critical need for a versatile, efficient tool that can bridge the gap between different genome alignment formats, thereby facilitating a more integrated approach to genomic analysis. Such a tool could enhance data compatibility and accessibility, and also enable more comprehensive and flexible analysis, fostering collaboration and innovation in genomic research. However, there is currently no integrated software offering this capability, which prevents researchers from fully capitalizing on the wealth of information in their datasets.

\begin{figure}[!t]%
\centering
\includegraphics[width=0.5\textwidth]{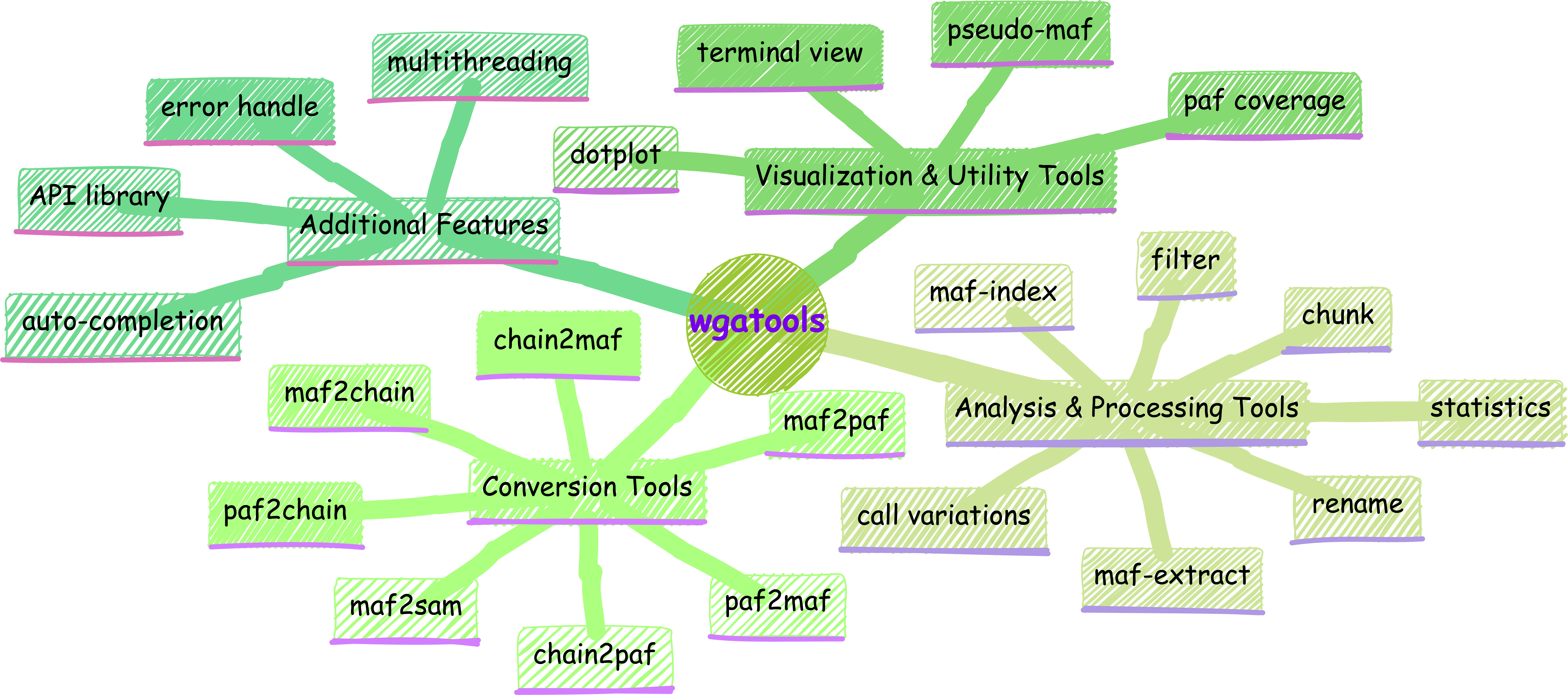}
\caption{
\textbf{Comprehensive suite of functionalities in \textit{wgatools}.}
}
\label{fig:funcs}
\end{figure}

Here, we have developed \textit{\textbf{wgatools}} to address this challenge. Programmed with Rust, \textit{\textbf{wgatools}} is an ultrafast, cross-platform toolkit designed to support all major WGA formats, thereby providing efficient conversion between them. \textit{\textbf{wgatools}} offers functionalities for processing, filtering, and statistically evaluating genome alignments, and includes features for variant calling and both local and genome-wide visualization (Figure~\ref{fig:funcs}). The toolkit performs efficiently on standard personal computers and is robust enough to handle large-scale genomic studies involving hundreds of genomes. \textit{\textbf{wgatools}} has already been used in newly developed multi-genome alignment pipeline \cite{zhou_acmga_2024}, in studying the evolution of complex regions that were previously uncharacterized \cite{Ape2024}, and is currently being integrated into more pan-genome pipelines.

\section{Implementation of wgatools}

\subsection{Format Conversion}
\textit{\textbf{wgatools}} is equipped with a variety of tools (Figure~\ref{fig:funcs}) to handle and transform genome alignment files across different formats, eliminating the need to start from scratch with specific workflows to generate particular formats. It supports conversion among three popular formats: MAF, PAF, and Chain. Our pivotal conversion step involves a set of byte-oriented, zero-copy, memory-safe, and exceptionally fast parsing combinators for the CIGAR string, an efficient compressed representation of alignment information. This ensures rapid and reliable parsing, significantly enhancing the performance and usability of our toolkit in genomic analysis.

\subsection{Data Processing and Analysis}
\textit{\textbf{wgatools}} offers extensive data processing and statistical analysis capabilities that significantly enhance its utility in genomic data analysis. The toolkit supports efficient indexing and precise extraction of specific intervals from MAF files, streamlining the handling of large alignment datasets. Furthermore, it allows for the segmentation of MAF files into smaller, sequential, and manageable chunks, facilitating parallel processing and subsequent analysis.

\textit{\textbf{wgatools}} provides comprehensive statistical summaries and filtering for various alignment files, offering valuable insights into alignment quality and characteristics. Additionally, it supports advanced functionalities for pairwise genome alignments across multiple species, including computing coverage metrics and generating pseudo-MAF format. These features are particularly useful for analyzing both single nucleotide polymorphism (SNPs) and gap divergence, as well as for studying conservation, making them integral to evolutionary research. Notably, \textit{\textbf{wgatools}} has been demonstrated its utility in investigating the evolution and diversity of previously uncharacterized and understudied genomic regions \cite{Ape2024}. By effectively managing and analyzing complex genomic data, \textit{\textbf{wgatools}} supports significant biological discoveries and enhances our understanding of genomic diversity.

\subsection{Variant Identification}
\textit{\textbf{wgatools}} employs efficient algorithms to identify various genomic variations by distinct alignment signatures, including SNPs, insertions, deletions, and other structural variations. By accurately detecting these mutations, researchers can gain valuable insights into the genetic diversity within and between species. The mutation identification module of \textit{\textbf{wgatools}} is highly customizable, allowing users to define specific output fields and filters to tailor the analysis to their research needs.

Owing to these advantages, \textit{\textbf{wgatools}} is being employed to evaluate the efficacy of state-of-the-art genome-wide alignment software (unpublished data), including wfmash \cite{wfmash_sftw}, minimap2 \cite{li_minimap2_2018}, and AnchorWave \cite{Song2021}. This application highlights the role of \textit{\textbf{wgatools}} as a reliable benchmark for comparing the performance of various alignment tools, reinforcing its utility in comprehensive genomic analysis and mutation detection.

\subsection{Visualization of Alignment Results}
Understanding complex genomic data can be greatly facilitated by effective visualization. \textit{\textbf{wgatools}} provides two visualization modules to help researchers explore and interpret genomic variations intuitively:
\begin{enumerate}
    \item \textbf{Terminal User Interface (TUI):} Given that most bioinformatics analyses are conducted through the terminal, this module is highly convenient. Users can execute commands and display results interactively using only the keyboard.
    \item \textbf{Interactive Dot Plot:} This module allows users to drag and scale the field of view freely, facilitating an in-depth understanding of the genomic relationships from various perspectives. Additionally, it offers the flexibility to switch between base-level and overview-level views, enhancing the interpretability of complex genomic data.
\end{enumerate}

\subsection{Performance and Usability}
\textit{\textbf{wgatools}} is written in Rust, a language known for its memory safety, concurrency support, and execution efficiency. This ensures robustness and efficient handling of large datasets. \textit{\textbf{wgatools}} stands out for its speed. Unlike the slower, disparate personal Python scripts often used in this domain, \textit{\textbf{wgatools}} is significantly faster, even when compared to similar Rust-based tools, in format conversion. For example, it achieves approximately five times faster performance than paf2chain \cite{andrea_guarracino_2023}.

Designed for user-friendly and versatility, it offers numerous parameters, shell auto-completion, robust error management, efficient multi-threading, and supports various compressed formats. In addition to its command-line tools, \textit{\textbf{wgatools}} also incorporates a comprehensive Rust library, providing developers with a powerful, low-level API for seamless integration into their software development processes and workflows. This facilitates efficient handling of genomic data and supports the development of high-performance, custom bioinformatics applications.

\textit{\textbf{wgatools}} is reproducible and reliable across multiple platforms. It can be easily installed from widely distributed sources like Bioconda, Nix, Docker, and Singularity.

\subsection{Future Directions}

Future development of \textit{\textbf{wgatools}} will focus on supporting more efficient formats such as HAL (Hierarchical ALignment), which is essential for comparative genomics \cite{Hickey_btt128}. Additionally, \textit{\textbf{wgatools}} will integrate formats related to graph-based pan-genomes, a key future direction in genomics. By supporting these advanced formats, \textit{\textbf{wgatools}} aims to facilitate comprehensive and ongoing genomic analysis continuously, ensuring it remains an essential tool for addressing the challenges posed by increasingly complex genomic datasets.

\section{Conclusions}
We have presented \textit{\textbf{wgatools}}, an ultrafast toolkit for manipulating whole genome alignments, representing a significant advancement in comparative genomics data analysis. \textit{\textbf{wgatools}} offers unprecedented speed and versatility, performing format conversion, data processing, statistical analysis, mutation identification, and visualization with efficiency. Its capabilities make it a valuable tool for researchers seeking to gain meaningful insights from complex genomic datasets.

\textit{\textbf{wgatools}} enhances downstream analyses by integrating data from different pipelines and swiftly converting between formats. It fosters collaboration and data sharing among researchers, enabling easy comparison and combination of results obtained using different methods. This would facilitate a deeper understanding of genomic variations and their impacts in various biological contexts.

The widespread adoption of \textit{\textbf{wgatools}} by numerous users highlights its utility and effectiveness, underscoring its reliability and performance as one of the leading tools for comparative genomic data analysis.

\section{Acknowledgements}
We thank the members of both Jianbing Yan's and Jian Yang's labs for their invaluable feedback and suggestions during the software development process. This project has received funding from the European Union's Framework Programme for Research and Innovation Horizon 2020 (2014-2020) under the Marie Curie Sk\l{}odowska Grant Agreement [Nr. 847548 to H.-J.L.], the ``Pioneer" and ``Leading Goose" R\&D Program of Zhejiang [2024SSYS0032 to J.Yang], and the NIH [R01HG013017 to E.G.] and NSF [\#2118743 to E.G.].

\section{Author Contributions}
Conceptualization, W.W. and S.-T.G.; Supervision, H.-J.L., J.Yan and J.Yang; Software, W.W.; Writing-Original Draft, H.-J.L. and W.W.; Writing-Review \& Editing, H.-J.L., W.W., J.Yang, J.Yan and E.G.

\section{Competing Interests}
The authors declare no competing interests.

\bibliographystyle{unsrt}
\bibliography{ref}

\begin{thebibliography}{10}

\bibitem{van_dijk_third_2018}
Erwin~L. Van~Dijk, Yan Jaszczyszyn, Delphine Naquin, and Claude Thermes.
\newblock The {Third} {Revolution} in {Sequencing} {Technology}.
\newblock {\em Trends in Genetics}, 34(9):666--681, September 2018.

\bibitem{li_genome_2024}
Heng Li and Richard Durbin.
\newblock Genome assembly in the telomere-to-telomere era.
\newblock {\em Nature Reviews Genetics}, April 2024.

\bibitem{anisimova_whole-genome_2019}
Colin~N. Dewey.
\newblock Whole-{Genome} {Alignment}.
\newblock In Maria Anisimova, editor, {\em Evolutionary {Genomics}}, volume 1910, pages 121--147. Springer New York, New York, NY, 2019.

\bibitem{kille_multiple_2022}
Bryce Kille, Advait Balaji, Fritz~J. Sedlazeck, Michael Nute, and Todd~J. Treangen.
\newblock Multiple genome alignment in the telomere-to-telomere assembly era.
\newblock {\em Genome Biology}, 23(1):182, August 2022.

\bibitem{song_new_2024}
Baoxing Song, Edward~S. Buckler, and Michelle~C. Stitzer.
\newblock New whole-genome alignment tools are needed for tapping into plant diversity.
\newblock {\em Trends in Plant Science}, 29(3):355--369, March 2024.

\bibitem{li_minimap2_2018}
Heng Li.
\newblock Minimap2: pairwise alignment for nucleotide sequences.
\newblock {\em Bioinformatics}, 34(18):3094--3100, September 2018.

\bibitem{zhou_acmga_2024}
Huafeng Zhou, Xiaoquan Su, and Baoxing Song.
\newblock {ACMGA}: a reference-free multiple-genome alignment pipeline for plant species.
\newblock {\em BMC Genomics}, 25(1):515, May 2024.

\bibitem{Ape2024}
DongAhn Yoo, Arang Rhie, Prajna Hebbar, Francesca Antonacci, Glennis~A. Logsdon, Steven~J. Solar, Dmitry Antipov, Brandon~D. Pickett, Yana Safonova, Francesco Montinaro, Yanting Luo, Joanna Malukiewicz, Jessica~M. Storer, Jiadong Lin, Abigail~N. Sequeira, Riley~J. Mangan, Glenn Hickey, Graciela~Monfort Anez, Parithi Balachandran, Anton Bankevich, Christine~R. Beck, Arjun Biddanda, Matthew Borchers, Gerard~G. Bouffard, Emry Brannan, Shelise~Y. Brooks, Lucia Carbone, Laura Carrel, Agnes~P. Chan, Juyun Crawford, Mark Diekhans, Eric Engelbrecht, Cedric Feschotte, Giulio Formenti, Gage~H. Garcia, Luciana~de Gennaro, David Gilbert, Richard~E. Green, Andrea Guarracino, Ishaan Gupta, Diana Haddad, Junmin Han, Robert~S. Harris, Gabrielle~A. Hartley, William~T. Harvey, Michael Hiller, Kendra Hoekzema, Marlys~L. Houck, Hyeonsoo Jeong, Kaivan Kamali, Manolis Kellis, Bryce Kille, Chul Lee, Youngho Lee, William Lees, Alexandra~P. Lewis, Qiuhui Li, Mark Loftus, Yong Hwee~Eddie Loh, Hailey Loucks, Jian Ma, Yafei Mao, Juan
  F.~I. Martinez, Patrick Masterson, Rajiv~C. McCoy, Barbara McGrath, Sean McKinney, Britta~S. Meyer, Karen~H. Miga, Saswat~K. Mohanty, Katherine~M. Munson, Karol Pal, Matt Pennell, Pavel~A. Pevzner, David Porubsky, Tamara Potapova, Francisca~R. Ringeling, Joana~L. Rocha, Oliver~A. Ryder, Samuel Sacco, Swati Saha, Takayo Sasaki, Michael~C. Schatz, Nicholas~J. Schork, Cole Shanks, Linnea Smeds, Dongmin~R. Son, Cynthia Steiner, Alexander~P. Sweeten, Michael~G. Tassia, Francoise Thibaud-Nissen, Edmundo Torres-Gonzalez, Mihir Trivedi, Wenjie Wei, Julie Wertz, Muyu Yang, Panpan Zhang, Shilong Zhang, Yang Zhang, Zhenmiao Zhang, Sarah~A. Zhao, Yixin Zhu, Erich~D. Jarvis, Jennifer~L. Gerton, Iker Rivas-Gonzalez, Benedict Paten, Zachary~A. Szpiech, Christian~D. Huber, Tobias~L. Lenz, Miriam~K. Konkel, Soojin~V. Yi, Stefan Canzar, Corey~T. Watson, Peter~H. Sudmant, Erin Molloy, Erik Garrison, Craig~B. Lowe, Mario Ventura, Rachel~J. O'Neill, Sergey Koren, Kateryna~D. Makova, Adam~M. Phillippy, and Evan~E. Eichler.
\newblock Complete sequencing of ape genomes, July 2024.
\newblock Pages: 2024.07.31.605654 Section: New Results.

\bibitem{wfmash_sftw}
Andrea Guarracino, Njagi Mwaniki, Santiago Marco-Sola, and Erik Garrison.
\newblock {wfmash: whole-chromosome pairwise alignment using the hierarchical wavefront algorithm}, June 2023.

\bibitem{Song2021}
Baoxing Song, Santiago Marco-Sola, Miquel Moreto, Lynn Johnson, Edward~S. Buckler, and Michelle~C. Stitzer.
\newblock Anchorwave: Sensitive alignment of genomes with high sequence diversity, extensive structural polymorphism, and whole-genome duplication.
\newblock {\em Proceedings of the National Academy of Sciences}, 119(1), December 2021.

\bibitem{andrea_guarracino_2023}
Andrea Guarracino.
\newblock Andreaguarracino/paf2chain: v0.1.0, July 2023.

\bibitem{Hickey_btt128}
Glenn Hickey, Benedict Paten, Dent Earl, Daniel Zerbino, and David Haussler.
\newblock {HAL: a hierarchical format for storing and analyzing multiple genome alignments}.
\newblock {\em Bioinformatics}, 29(10):1341--1342, 03 2013.

\end{thebibliography}


\end{document}